\documentclass[final]{cvpr}

\usepackage{times}
\usepackage{epsfig}
\usepackage{graphicx}
\usepackage{amsmath}
\usepackage{amssymb}
\usepackage[table,x11names]{xcolor}

\usepackage[pagebackref=true,breaklinks=true,colorlinks,bookmarks=false]{hyperref}

\def\cvprPaperID{****} 
\def\confYear{CVPR 2021}
\definecolor{grayhighlight}{RGB}{213,229,255}

\newsavebox\CBox
\def\textBF#1{\sbox\CBox{#1}\resizebox{\wd\CBox}{\ht\CBox}{\textbf{#1}}}

\begin{document}

\title{Real-Time Video Super-Resolution on Smartphones with Deep Learning,\\ Mobile AI 2021 Challenge: Report}
\author{
Andrey Ignatov \and Andres Romero \and Heewon Kim \and Radu Timofte \and
Chiu Man Ho \and Zibo Meng \and
Kyoung Mu Lee \and
Yuxiang Chen \and Yutong Wang \and Zeyu Long \and Chenhao Wang \and Yifei Chen \and Boshen Xu \and Shuhang Gu \and Lixin Duan \and Wen Li \and
Wang Bofei \and Zhang Diankai \and Zheng Chengjian \and Liu Shaoli \and Gao Si \and Zhang xiaofeng \and Lu Kaidi \and Xu Tianyu \and
Zheng Hui \and Xinbo Gao \and Xiumei Wang \and
Jiaming Guo \and Xueyi Zhou \and Hao Jia \and Youliang Yan
}

\maketitle

\begin{abstract}

Video super-resolution has recently become one of the most important mobile-related problems due to the rise of video communication and streaming services. While many solutions have been proposed for this task, the majority of them are too computationally expensive to run on portable devices with limited hardware resources. To address this problem, we introduce the first Mobile AI challenge, where the target is to develop an end-to-end deep learning-based video super-resolution solutions that can achieve a real-time performance on mobile GPUs. The participants were provided with the REDS dataset and trained their models to do an efficient 4X video upscaling. The runtime of all models was evaluated on the OPPO Find X2 smartphone with the Snapdragon 865 SoC capable of accelerating floating-point networks on its Adreno GPU. The proposed solutions are fully compatible with any mobile GPU and can upscale videos to HD resolution at up to 80 FPS while demonstrating high fidelity results. A detailed description of all models developed in the challenge is provided in this paper.

\end{abstract}
{\let\thefootnote\relax\footnotetext{%
\hspace{-5mm}$^*$Andrey Ignatov, Andres Romero, Heewon Kim and Radu Timofte \textit{(andrey@vision.ee.ethz.ch, roandres@ethz.ch, ghimhw@gmail.com, radu.timofte@vision.ee.ethz.ch)} are the main Mobile AI 2021 challenge organizers. The other authors participated in the challenge. \\ Appendix \ref{sec:apd:team} contains the authors' team names and affiliations. \vspace{2mm} \\ Mobile AI 2021 Workshop website: \\ \url{https://ai-benchmark.com/workshops/mai/2021/}
}}

\section{Introduction}

\begin{figure*}[t!]
\centering
\setlength{\tabcolsep}{1pt}
\resizebox{\linewidth}{!}
{
\begin{tabular}{cccccc}
Bicubic Upscaling & Team ZTE VIP & Team Noah\_TerminalVision & Team Rainbow & Team Diggers & High-Resolution Target \\
\\
\includegraphics[width=0.3\linewidth]{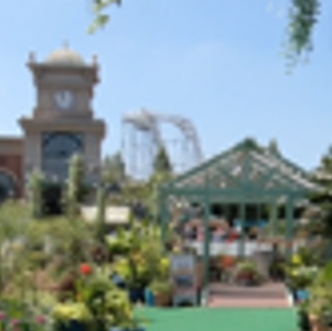}&
\includegraphics[width=0.3\linewidth]{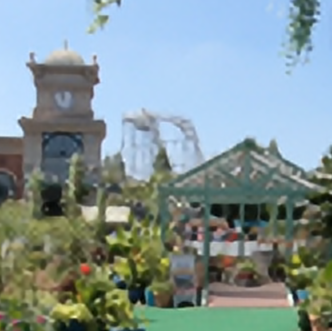}&
\includegraphics[width=0.3\linewidth]{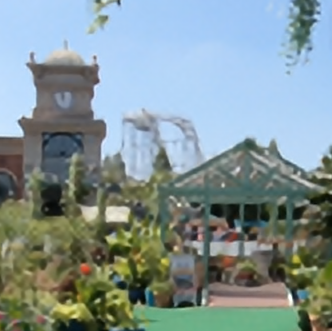}&
\includegraphics[width=0.3\linewidth]{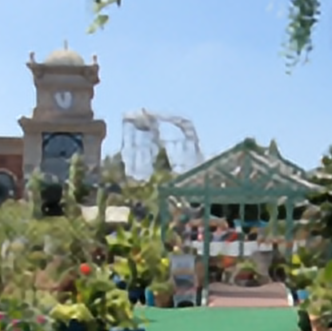}&
\includegraphics[width=0.3\linewidth]{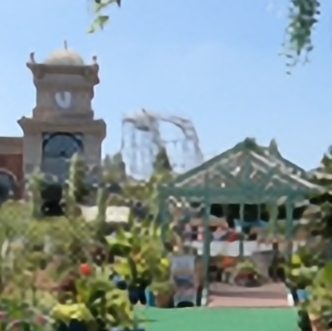}&
\includegraphics[width=0.3\linewidth]{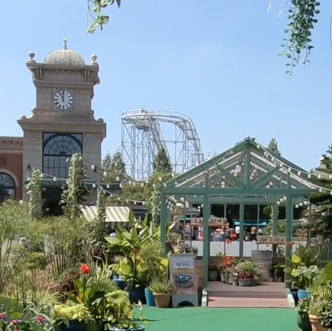}\\
\includegraphics[width=0.3\linewidth]{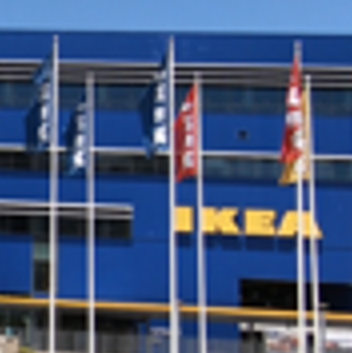}&
\includegraphics[width=0.3\linewidth]{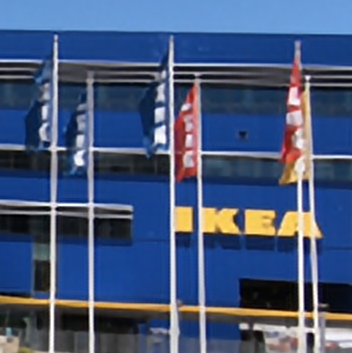}&
\includegraphics[width=0.3\linewidth]{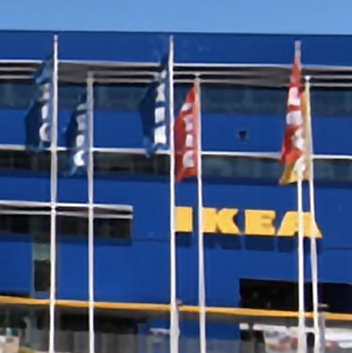}&
\includegraphics[width=0.3\linewidth]{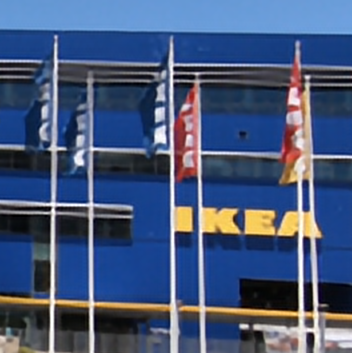}&
\includegraphics[width=0.3\linewidth]{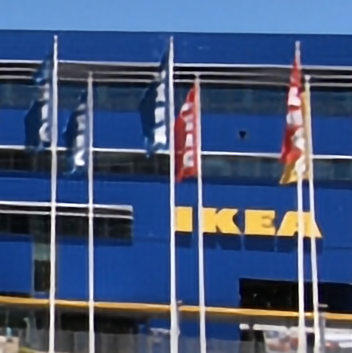}&
\includegraphics[width=0.3\linewidth]{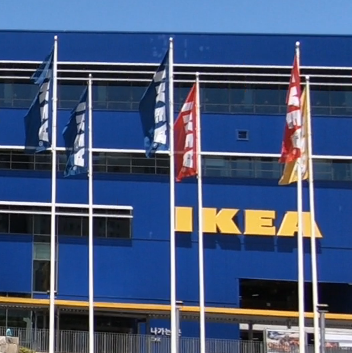}\\
\includegraphics[width=0.3\linewidth]{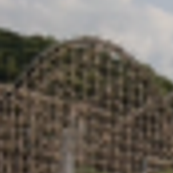}&
\includegraphics[width=0.3\linewidth]{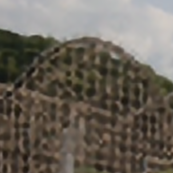}&
\includegraphics[width=0.3\linewidth]{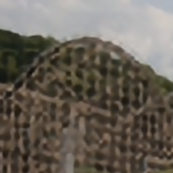}&
\includegraphics[width=0.3\linewidth]{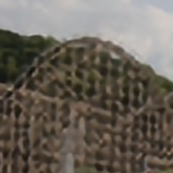}&
\includegraphics[width=0.3\linewidth]{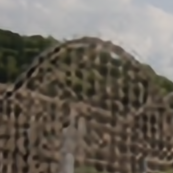}&
\includegraphics[width=0.3\linewidth]{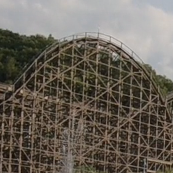}\\
\includegraphics[width=0.3\linewidth]{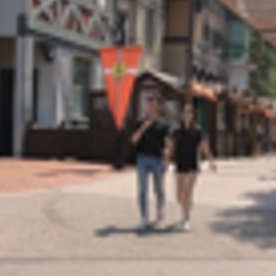}&
\includegraphics[width=0.3\linewidth]{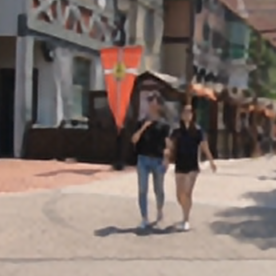}&
\includegraphics[width=0.3\linewidth]{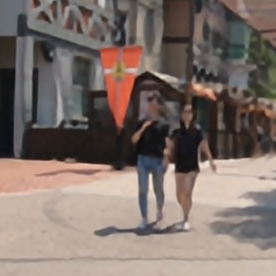}&
\includegraphics[width=0.3\linewidth]{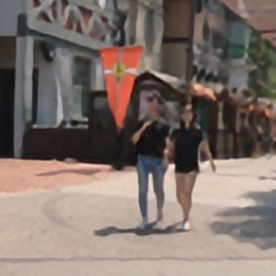}&
\includegraphics[width=0.3\linewidth]{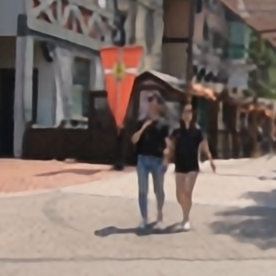}&
\includegraphics[width=0.3\linewidth]{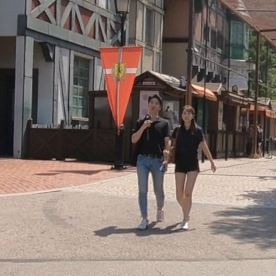}\\
\end{tabular}
}
\vspace{0cm}
\caption{Sample crops from the original video frames, results obtained by challenge participants and the target high-resolution frames.}
\label{fig:qualitative}
\vspace{-0.2cm}
\end{figure*}

An increased popularity of various video streaming services and a widespread of mobile devices have created a strong need for efficient and mobile-friendly video super-resolution approaches. Over the past years, many accurate deep learning-based solutions have been proposed for this problem~\cite{nah2019ntire,nah2019ntireChallenge,wang2019edvr,kappeler2016video,shi2016real,sajjadi2018frame,fuoli2019efficient}. The biggest limitation of these methods is that they were primarily targeted at achieving high fidelity scores while not optimized for computational efficiency and mobile-related constraints, which is essential for tasks related to image~\cite{ignatov2017dslr,ignatov2018wespe,ignatov2020replacing} and video~\cite{nah2020ntire} enhancement on mobile devices. In this challenge, we take one step further in solving this problem by using a popular REDS~\cite{nah2019ntire} video super-resolution dataset and by imposing additional efficiency-related constraints on the developed solutions.

When it comes to the deployment of AI-based solutions on mobile devices, one needs to take care of the particularities of mobile NPUs and DSPs to design an efficient model. An extensive overview of smartphone AI acceleration hardware and its performance is provided in~\cite{ignatov2019ai,ignatov2018ai}. According to the results reported in these papers, the latest mobile NPUs are already approaching the results of mid-range desktop GPUs released not long ago. However, there are still two major issues that prevent a straightforward deployment of neural networks on mobile devices: a restricted amount of RAM, and a limited and not always efficient support for many common deep learning layers and operators. These two problems make it impossible to process high resolution data with standard NN models, thus requiring a careful adaptation of each architecture to the restrictions of mobile AI hardware. Such optimizations can include network pruning and compression~\cite{chiang2020deploying,ignatov2020rendering,li2019learning,liu2019metapruning,obukhov2020t}, 16-bit / 8-bit~\cite{chiang2020deploying,jain2019trained,jacob2018quantization,yang2019quantization} and low-bit~\cite{cai2020zeroq,uhlich2019mixed,ignatov2020controlling,liu2018bi} quantization, device- or NPU-specific adaptations, platform-aware neural architecture search~\cite{howard2019searching,tan2019mnasnet,wu2019fbnet,wan2020fbnetv2}, \etc.

While many challenges and works targeted at efficient deep learning models have been proposed recently, the evaluation of the obtained solutions is generally performed on desktop CPUs and GPUs, making the developed solutions not practical due to the above mentioned issues. To address this problem, we introduce the first \textit{Mobile AI Workshop and Challenges}, where all deep learning solutions are developed for and evaluated on real mobile devices.
In this competition, the participating teams were provided with the original high-quality and downscaled by a factor of 4 videos from the REDS~\cite{nah2019ntire} dataset to train their networks. Within the challenge, the participants were evaluating the runtime and tuning their models on the OPPO Find X2 smartphone featuring the Qualcomm Adreno 650 GPU that can efficiently accelerate floating-point neural networks.
The final score of each submitted solution was based on the runtime and fidelity results, thus balancing between the image reconstruction quality and efficiency of the proposed model. Finally, all developed solutions are fully compatible with the TensorFlow Lite framework~\cite{TensorFlowLite2021}, thus can be deployed and accelerated on any mobile platform providing AI acceleration through the Android Neural Networks API (NNAPI)~\cite{NNAPI2021} or custom TFLite delegates~\cite{TFLiteDelegates2021}.

\smallskip


This challenge is a part of the \textit{MAI 2021 Workshop and Challenges} consisting of the following competitions:


\small

\begin{itemize}
\item Learned Smartphone ISP on Mobile NPUs~\cite{ignatov2021learned}
\item Real Image Denoising on Mobile GPUs~\cite{ignatov2021fastDenoising}
\item Quantized Image Super-Resolution on Mobile NPUs~\cite{ignatov2021real}
\item Real-Time Video Super-Resolution on Mobile GPUs
\item Single-Image Depth Estimation on Mobile Devices~\cite{ignatov2021fastDepth}
\item Quantized Camera Scene Detection on Smartphones~\cite{ignatov2021fastSceneDetection}
\item High Dynamic Range Image Processing on Mobile NPUs
\end{itemize}

\normalsize


\noindent The results obtained in the other competitions and the description of the proposed solutions can be found in the corresponding challenge papers.


\begin{figure*}[t!]
\centering
\setlength{\tabcolsep}{1pt}
\resizebox{0.96\linewidth}{!}
{
\includegraphics[width=1.0\linewidth]{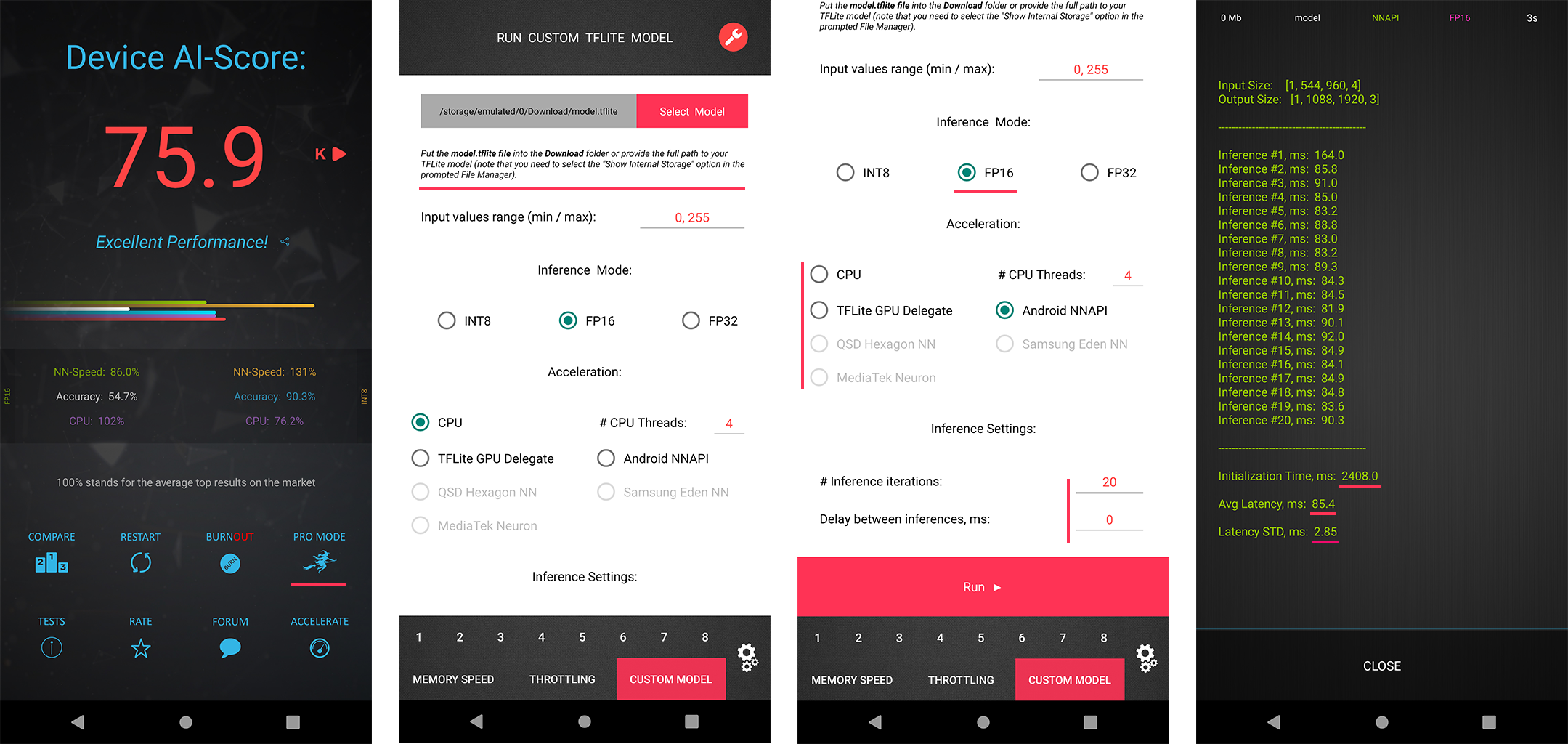}
}
\vspace{0.2cm}
\caption{Loading and running custom TensorFlow Lite models with AI Benchmark application. The currently supported acceleration options include Android NNAPI, TFLite GPU, Hexagon NN, Samsung Eden and MediaTek Neuron delegates as well as CPU inference through TFLite or XNNPACK backends. The latest app version can be downloaded at \url{https://ai-benchmark.com/download}}
\vspace{-0.2cm}
\label{fig:ai_benchmark_custom}
\end{figure*}

\section{Challenge}

To develop an efficient and practical solution for mobile-related tasks, one needs the following major components:

\begin{enumerate}
\item A high-quality and large-scale dataset that can be used to train and evaluate the solution;
\item An efficient way to check the runtime and debug the model locally without any constraints;
\item An ability to regularly test the runtime of the designed neural network on the target mobile platform or device.
\end{enumerate}

This challenge addresses all the above issues. Real training data, tools, and runtime evaluation options provided to the challenge participants are described in the next sections.

\subsection{Dataset}

In this challenge, we use the REDS~\cite{nah2019ntire} dataset that serves as a benchmark for traditional video super-resolution task as it contains a large diversity of content and dynamic scenes. Following the standard procedure, we use 240 videos for training, 30 videos for validation, and 30 videos for testing. Each video has sequences of length 100, where every sequence contains video frames of 1280$\times$720 resolution at 24 fps. To generate low-resolution data, the videos were bicubically downsampled with a factor of 4. The low-resolution video data is then considered as input, and the high-resolution~--- are the target.

\begin{table*}[t!]
\centering
\resizebox{\linewidth}{!}
{
\begin{tabular}{l|c|cc|cc|ccc|c}
\hline
Team \, & \, Author \, & \, Framework \, & Model Size, & \, PSNR$\uparrow$ \, & \, SSIM$\uparrow$ \, & \multicolumn{2}{c}{\, Runtime per 10 frames $\downarrow$ \,} & Speed-Up & \, Final Score \\
& & & KB & & & \, CPU, ms \, & \, GPU, ms \, & \\
\hline
\hline
Diggers & \, chenyuxiang \, & Keras / TensorFlow & 230 & \textBF{28.33} & \textBF{0.8112} & 916 & 199 & 4.6 & \textBF{8.13} \\
ZTE VIP & jieson\_zheng & \, PyTorch / TensorFlow \, & 50 & 27.85 & 0.7983 & 163 & \textBF{113} & 1.4 & 7.36 \\
Rainbow & Zheng222 & TensorFlow & 204 & 27.99 & 0.8021 & 429 & 180 & 2.4 & 5.61 \\
Noah\_TerminalVision \, & JeremieG & TensorFlow & 30 & 27.97 & 0.8017 & 448 & \footnotesize{\textcolor{red}{Error $^*$}} & - & 2.19 \\
\rowcolor{grayhighlight} Bicubic Upscaling & Baseline & & & 26.50 & 0.7508 & & & & - \\
\end{tabular}
}
\vspace{2.6mm}
\caption{\small{Mobile AI 2021 Real-Time Video Super-Resolution challenge results and final rankings. During the runtime measurements, the models were upscaling 10 subsequent video frames from 180$\times$320 to 1280$\times$720 pixels on the OPPO Find X2 smartphone. Team \textit{Diggers} is the challenge winner. $^*$~The solution from \textit{Noah\_TerminalVision} was not parsed correctly by the TFLite GPU delegate.}}
\label{tab:results}
\end{table*}

\subsection{Local Runtime Evaluation}

When developing AI solutions for mobile devices, it is vital to be able to test the designed models and debug all emerging issues locally on available devices. For this, the participants were provided with the \textit{AI Benchmark} application~\cite{ignatov2018ai,ignatov2019ai} that allows to load any custom TensorFlow Lite model and run it on any Android device with all supported acceleration options. This tool contains the latest versions of \textit{Android NNAPI, TFLite GPU, Hexagon NN, Samsung Eden} and \textit{MediaTek Neuron} delegates, therefore supporting all current mobile platforms and providing the users with the ability to execute neural networks on smartphone NPUs, APUs, DSPs, GPUs and CPUs.

\smallskip

To load and run a custom TensorFlow Lite model, one needs to follow the next steps:

\begin{enumerate}
\setlength\itemsep{0mm}
\item Download AI Benchmark from the official website\footnote{\url{https://ai-benchmark.com/download}} or from the Google Play\footnote{\url{https://play.google.com/store/apps/details?id=org.benchmark.demo}} and run its standard tests.
\item After the end of the tests, enter the \textit{PRO Mode} and select the \textit{Custom Model} tab there.
\item Rename the exported TFLite model to \textit{model.tflite} and put it into the \textit{Download} folder of the device.
\item Select mode type \textit{(INT8, FP16, or FP32)}, the desired acceleration/inference options and run the model.
\end{enumerate}

\noindent These steps are also illustrated in Fig.~\ref{fig:ai_benchmark_custom}.

\subsection{Runtime Evaluation on the Target Platform}

In this challenge, we use the \textit{OPPO Find X2} smartphone with the \textit{Qualcomm Snapdragon 865} mobile SoC as our target runtime evaluation platform. The considered chipset demonstrates very decent AI Benchmark scores~\cite{AIBenchmark202104} and can be found in the majority of flagship Android smartphones released in 2020. It can efficiently accelerate floating-point networks on its Adreno 650 GPU with a theoretical FP16 performance of 2.4 TFLOPS. Within the challenge, the participants were able to upload their TFLite models to an external server and get a feedback regarding the speed of their model: the runtime of their solution on the above mentioned OPPO device or an error log if the network contains some incompatible operations. The models were parsed and accelerated using the TensorFlow Lite GPU delegate~\cite{lee2019device} demonstrating the best performance on this platform according to AI Benchmark results. The same setup was also used for the final runtime evaluation.

\subsection{Challenge Phases}

The challenge consisted of the following phases:

\vspace{-0.8mm}
\begin{enumerate}
\item[I.] \textit{Development:} the participants get access to the data and AI Benchmark app, and are able to train the models and evaluate their runtime locally;
\item[II.] \textit{Validation:} the participants can upload their models to the remote server to check the fidelity scores on the validation dataset, to get the runtime on the target platform, and to compare their results on the validation leaderboard;
\item[III.] \textit{Testing:} the participants submit their final results, codes, TensorFlow Lite models, and factsheets.
\end{enumerate}
\vspace{-0.8mm}

\subsection{Scoring System}

All solutions were evaluated using the following metrics:

\vspace{-0.8mm}
\begin{itemize}
\setlength\itemsep{-0.2mm}
\item Peak Signal-to-Noise Ratio (PSNR) measuring fidelity score,
\item Structural Similarity Index Measure (SSIM), a proxy for perceptual score,
\item The runtime on the target OPPO Find X2 smartphone.
\end{itemize}
\vspace{-0.8mm}

The goal of this challenge was to produce an efficient solution balancing between the fidelity scores and latency. For the fidelity evaluation, we compute the PSNR and SSIM measures between the target sharp high-resolution and the produced super-resolved videos, both scores were averaged over the entire sequence of frames. Different to common VSR methods~\cite{wang2019edvr,chan2020basicvsr} where the input of the model is a 5-dimensional tensor including the video sequence information, this challenge encouraged the participants to build models that receive mobile-friendly 4-dimensional tensors.
The input model tensor should accept 10 subsequent video frames and have a size of [$1 \times 180 \times 320 \times 30$], where the first dimension is the batch size, the second and third dimensions are the height and width of the input frames, and the last dimension is the number of channels (3 color channels $\times$ 10 frames). The size of the output tensor of the model should be [$1 \times 720 \times 1280 \times 30$].

The score of each final submission was evaluated based on the next formula ($C$ is a constant normalization factor):

\smallskip
\begin{equation*}
\text{Final Score} \,=\, \frac{2^{2 \cdot \text{PSNR}}}{C \cdot \text{runtime}},
\end{equation*}
\smallskip

During the final challenge phase, the participants did not have access to the test dataset. Instead, they had to submit their final TensorFlow Lite models that were subsequently used by the challenge organizers to check both the runtime and the fidelity results of each submission under identical conditions. This approach solved all the issues related to model overfitting, reproducibility of the results, and consistency of the obtained runtime/accuracy values.

\begin{figure*}[t!]
\centering
\resizebox{1.0\linewidth}{!}
{
\includegraphics[width=1.0\linewidth]{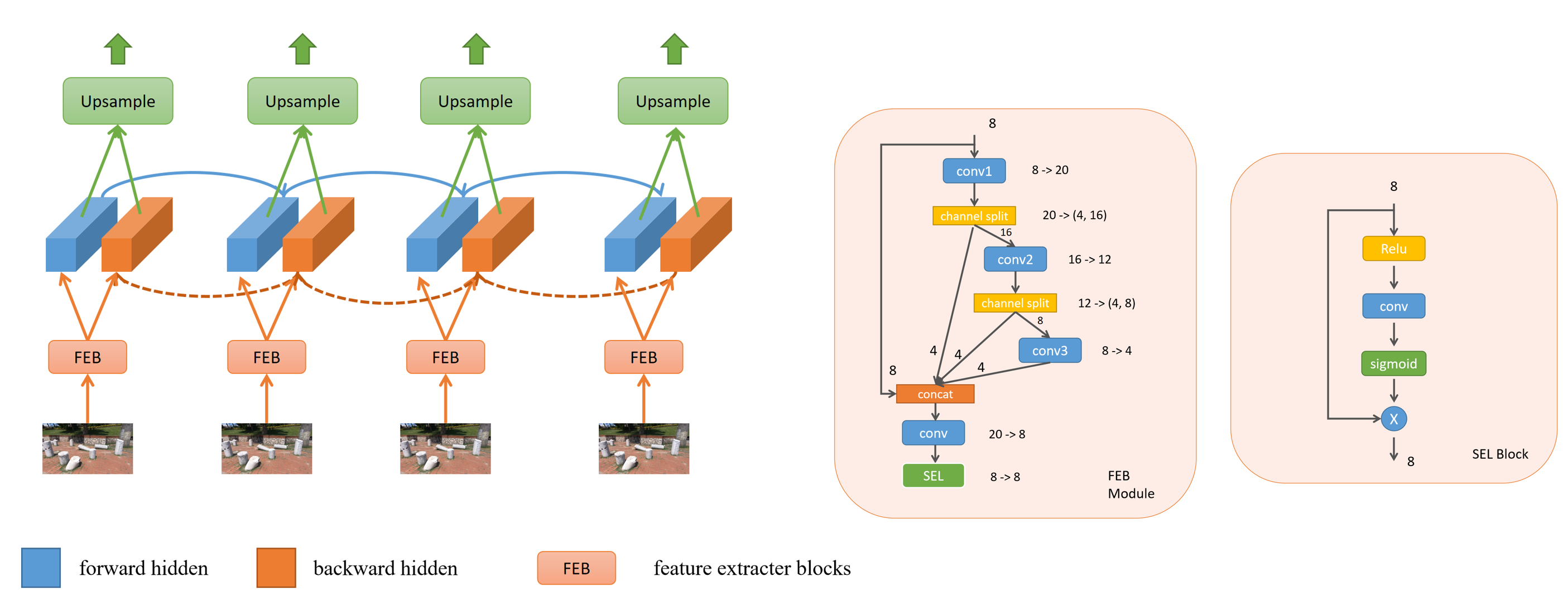}
}
\caption{\small{Team Diggers proposes a bidirectional RNN with efficient feature extractors (FEB) to exploit the temporal dependencies.}}
\label{fig:Diggers}
\end{figure*}

\section{Challenge Results}

From above 125 registered participants, 4 teams entered the final phase and submitted valid results, TFLite models, codes, executables and factsheets. Table~\ref{tab:results} summarizes the final challenge results and reports PSNR, SSIM and runtime numbers for each submitted solution on the final test dataset and on the target evaluation platform, while Fig.~\ref{fig:qualitative} shows the obtained qualitative results. The proposed methods are described in section~\ref{sec:solutions}, and the team members and affiliations are listed in Appendix~\ref{sec:apd:team}.

\subsection{Results and Discussion}

All submitted solutions demonstrated a very high efficiency: the first three models can upscale video frames from 180$\times$320 to 1280$\times$720 resolution at more than 50 FPS on the target Snapdragon 865 chipset. The solution proposed by team \textit{Noah\_TerminalVision} can potentially achieve even higher frame rates, though it is currently not compatible with the TFLite GPU delegate due to \textit{split} operations and thus was tested on Snapdragon's CPU only. Team \textit{Diggers} is the challenge winner~--- the model proposed by the authors achieves the best fidelity results while demonstrating good runtime values. This is the only solution in this challenge that applied recurrent connections to make use of inter-frame dependencies for getting better reconstruction results. While the other methods were performing only a standard per-frame upscaling, the image crops shown in Fig.~\ref{fig:qualitative} demonstrate that the visual quality of the reconstructed video frames obtained in their cases is just slightly behind the one of the winning solution, while all results are significantly better compared to the simple bicubic video interpolation. Therefore, we can conclude that, from the practical aspect, all proposed solutions can be applied for video super-resolution task on real mobile devices~--- the final choice will depend on the set of supported ops (\eg, the first model might not be compatible with some mobile NPUs~\cite{ignatov2021real}) and the target FPS and runtime values.

\section{Challenge Methods}
\label{sec:solutions}

\noindent This section describes solutions submitted by all teams participating in the final stage of the MAI 2021 Real-Time Video Super-Resolution challenge.

\subsection{Diggers}
\label{sec:diggers}

Team Diggers proposed a bi-directional recurrent model for the considered video super-resolution task that uses feature maps computed for the previous and future video frames as an additional information while super-resolving the current frame (Fig.~\ref{fig:Diggers}). The model architecture is generally based on the ideas proposed in~\cite{isobe2020revisiting} and~\cite{hui2019lightweight}: for each input frame, two feature extraction blocks (FEBs) are applied to generate the corresponding feature maps: forward [blue] and backward [orange]. The forward feature maps of the current and previous frames are then combined and passed to another feature extraction block to generate the final forward feature map for the current frame. As for the backward frames, the procedure is exactly the same, though the sequence is reversed. The obtained final forward and backward features are fed to the selection units layer (SEL) module~\cite{choi2017deep}, one IMDB~\cite{hui2019lightweight} module and two convolutional and image resizing layers performing final frame upscaling.

During the training process, sequences of 21 subsequent low- and high-resolution video frames were used as model's inputs and targets. First, the model was trained for 31 epochs with a batch size of 16 and an initial learning rate of $4e-3$ multiplied by $0.7$ each 2 epochs starting from the 7th one. Next, it was trained for another 31 epochs with a batch size of 32 and the same learning rate policy. $L_2$ loss was used as a target loss function, model parameters were optimized with Adam~\cite{kingma2014adam}. The images were additionally flipped randomly during the training for data augmentation.

\subsection{ZTE VIP}
\label{sec:zte}

\begin{figure}[h!]
\centering
\resizebox{1.0\linewidth}{!}
{
\includegraphics[width=1.0\linewidth]{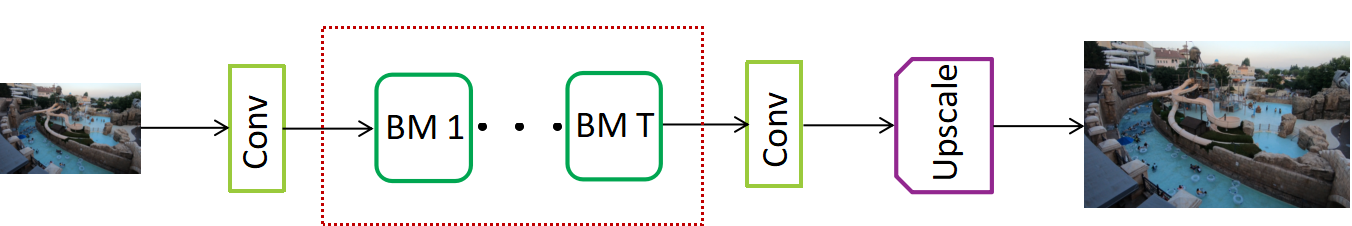}
}
\caption{\small{Feed-forward CNN proposed by ZTE VIP team. Each basic module (BM) consists of a residual block with two convolutional layers.}}
\label{fig:ZTE}
\end{figure}

Team ZTE VIP proposed a model performing per-frame upscaling without taking into account any inter-frame dependencies (Fig.~\ref{fig:ZTE}) which can significantly speed-up the inference. In its first layer, the input is resized so that the batch size is equal to the number of video frames, and then they are processed separately by several residual blocks~\cite{he2016deep} and a \textit{depth-to-space} layer performing final frame upsampling. The number of residual blocks and their size was found using the Neural Architecture Search (NAS)~\cite{kim2019fine}, where the target metric was composed of the fidelity loss and the number of model FLOPS. The final model contains five residual blocks, each one consisting of two 3$\times$3 convolutions with eight feature maps. The network was trained to minimize $L_1$ loss with a batch size of 4 for 1000 epochs using Adam optimizer with a learning rate of $2e-4$ down-scaled by a factor of 0.5 till 400 epochs. A more detailed description of the model, design choices and training procedure is provided in~\cite{liu2021EVSRNet}.

\subsection{Rainbow}
\label{sec:rainbow}

\begin{figure}[h!]
\centering
\resizebox{1.0\linewidth}{!}
{
\includegraphics[width=0.22\linewidth]{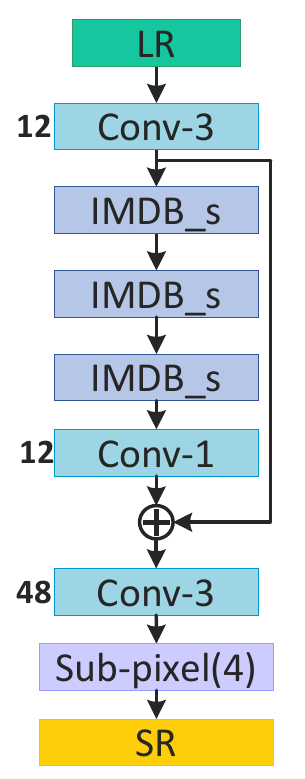}
\includegraphics[width=0.5\linewidth]{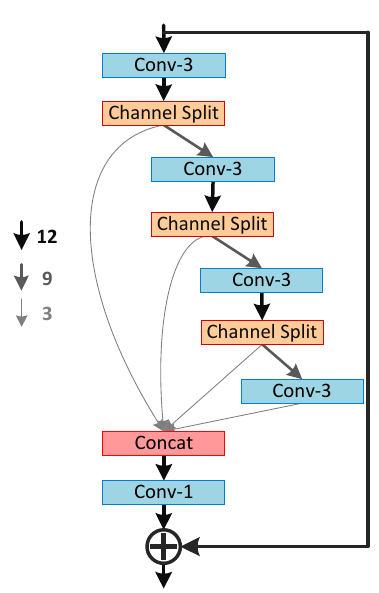}
}
\caption{\small{The overall model architecture proposed by Rainbow team (left). The structure of the information multi-distillation blocks (IMDB\_s) is presented on the right side; 12, 9 and 3 denote the size of the output channels of each convolutional layer.}}
\label{fig:Rainbow}
\end{figure}

Similarly to the previous solution, Rainbow team has developed a pure CNN model performing per-frame video upscaling (Fig.~\ref{fig:Rainbow}). The authors presented a network consisting of three information multi-distillation blocks (IMDB\_s)~\cite{hui2019lightweight} followed by a \textit{depth-to-space} upsampling layer processing each video frame separately. A global skip connection is used to improve the fidelity results of the model. $L_1$ loss was used as a target fidelity measure, network parameters were optimized using Adam with an initial learning rate of $2e-4$ halved every 50K iterations. A batch size of 8 was used during the training, horizontal and vertical flipping was used to augment the training data.

\subsection{Noah\_TerminalVision}
\label{sec:noah}

\begin{figure}[h!]
\centering
\resizebox{1.0\linewidth}{!}
{
\includegraphics[width=\linewidth]{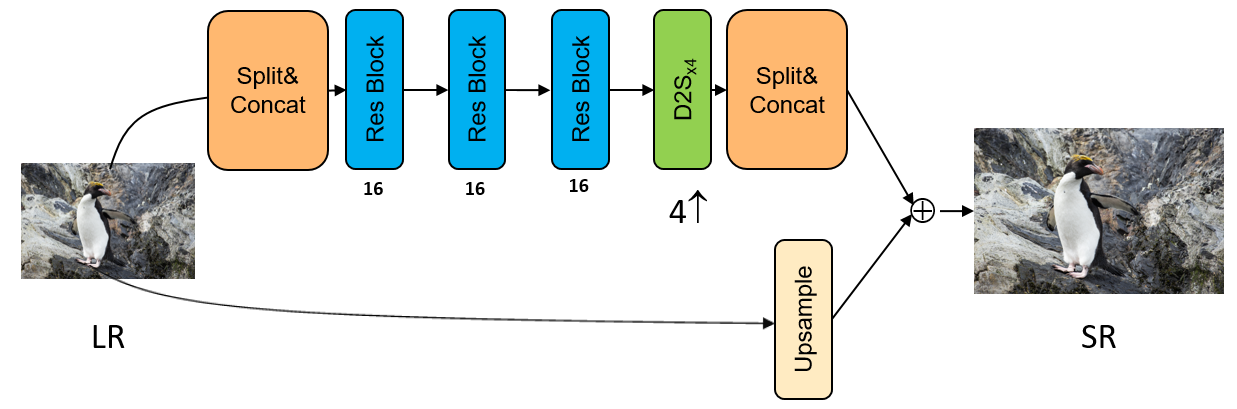}
}
\caption{\small{The solution proposed by Noah\_TerminalVision consists of a light-weight architecture with three residual blocks and asymmetric  convolutions.}}
\label{fig:Noah}
\end{figure}

Team Noah\_TerminalVision presented a TinyVSRNet architecture that contains three residual blocks (each consisting of 2 convolutions with 16 channels) followed by \textit{depth-to-space} upsampling layer and one global skip connection performing bilinear image upscaling (Fig.~\ref{fig:Noah}). The authors have also proposed a ``single-frame'' solution by converting 10 video frames from the channel dimension to the batch dimension with \textit{split} and \textit{concat} layers. To boost the model performance, they adopted the approach developed in~\cite{ding2019acnet}, where three asymmetric convolution kernels (of size 3$\times$3, 1$\times$3, and 3$\times$1) are used during the training and then fused into one single convolution op during the inference. With this modification, the results of the TinyVSRNet were improved by around 0.05 dB. The network was trained to minimize $L_1$ loss, its parameters were optimized using Adam for one million iterations with a cyclic learning rate starting from $5e-4$ and decreased to $1e-6$ each 200K iterations.

\section{Additional Literature}

An overview of the past challenges on mobile-related tasks together with the proposed solutions can be found in the following papers:

\begin{itemize}
\item Video Super-Resolution:\, \cite{nah2019ntireChallenge,nah2020ntire}
\item Image Super-Resolution:\, \cite{ignatov2018pirm,lugmayr2020ntire,cai2019ntire,timofte2018ntire}
\item Learned End-to-End ISP:\, \cite{ignatov2019aim,ignatov2020aim}
\item Perceptual Image Enhancement:\, \cite{ignatov2018pirm,ignatov2019ntire}
\item Bokeh Effect Rendering:\, \cite{ignatov2019aimBokeh,ignatov2020aimBokeh}
\item Image Denoising:\, \cite{abdelhamed2020ntire,abdelhamed2019ntire}
\end{itemize}

\section*{Acknowledgements}

We thank OPPO Mobile Co., Ltd., AI Witchlabs, ETH Zurich (Computer Vision Lab) and Seoul National University (SNU), the organizers and sponsors of this Mobile AI 2021 challenge.

\appendix
\section{Teams and Affiliations}
\label{sec:apd:team}

\bigskip

\subsection*{Mobile AI 2021 Team}
\noindent\textit{\textbf{Title: }}\\ Mobile AI 2021 Image Super-Resolution Challenge\\
\noindent\textit{\textbf{Members:}}\\ Andrey Ignatov$^{1,3}$ \textit{(andrey@vision.ee.ethz.ch)}, Andres Romero$^1$ \textit{(roandres@ethz.ch)}, Heewon  Kim$^4$ \textit{(ghimhw@gmail.com)}, Radu Timofte$^{1,3}$  \textit{(radu.timofte @vision.ee.ethz.ch)}, Chiu Man Ho$^2$ \textit{(chiuman@oppo.com)}, Zibo Meng$^2$ \textit{(zibo.meng@oppo.com)},
Kyoung Mu Lee$^4$\\
\noindent\textit{\textbf{Affiliations: }}\\
$^1$ Computer Vision Lab, ETH Zurich, Switzerland\\
$^2$ OPPO Mobile Co., Ltd, China\\
$^3$ AI Witchlabs, Switzerland\\
$^4$ Seoul National University, South Korea\\

\subsection*{Diggers}
\noindent\textit{\textbf{Title:}}\\ Real-Time Video Super-Resolution based on Bidirectional RNNs\\
\noindent\textit{\textbf{Members: }}\\ \textit{Yuxiang Chen (chenyx.cs@gmail.com)}, Yutong Wang, Zeyu Long, Chenhao Wang, Yifei Chen, Boshen Xu, Shuhang Gu, Lixin Duan, Wen Li\\
\noindent\textit{\textbf{Affiliations: }}\\
University of Electronic Science and Technology of China (UESTC), China\\

\subsection*{ZTE VIP}
\noindent\textit{\textbf{Title:}}\\ Evsrnet: Efficient video super-resolution with neural architecture search.~\cite{liu2021EVSRNet} \\
\noindent\textit{\textbf{Members:}}\\ \textit{Wang Bofei (wang.bofei@zte.com.cn)}, Zhang Diankai, Zheng Chengjian,
Liu Shaoli, Gao Si, Zhang xiaofeng, Lu Kaidi, Xu Tianyu\\
\noindent\textit{\textbf{Affiliations: }}\\
Audio \& Video Technology Platform Department, ZTE Co., Ltd, China\\

\subsection*{Rainbow}
\noindent\textit{\textbf{Title:}}\\Video Super-Resolution with Multi-Information Distillation Network on Mobile Device\\
\noindent\textit{\textbf{Members:}}\\ \textit{Zheng Hui (zheng\_hui@aliyun.com)}, Xinbo Gao, Xiumei Wang\\
\noindent\textit{\textbf{Affiliations: }}\\
School of Electronic Engineering, Xidian University, China\\

\subsection*{Noah\_TerminalVision}
\noindent\textit{\textbf{Title:}}\\TinyVSRNet for Real-Time Video Super-Resolution\\
\noindent\textit{\textbf{Members:}}\\\textit{Jiaming Guo (guojiaming5@huawei.com)}, Xueyi Zhou, Hao Jia, Youliang Yan\\
\noindent\textit{\textbf{Affiliations: }}\\
Huawei Technologies Co., Ltd, China\\

{\small
\bibliographystyle{ieee_fullname}

}

\end{document}